\documentclass[nofootinbib,longbibliography,prd,floatfix,superscriptaddress,11pt,tightenlines]{revtex4-1}
\usepackage[english]{babel}
\usepackage{amsmath,amssymb,amsbsy,amstext,amsthm,booktabs,simplewick,exscale,relsize,graphicx,amsfonts,upgreek}
\usepackage{multirow,dcolumn,bm,enumerate,url,hyperref}

\newcommand{\tr}{\text{track}}
\newcommand{\lsp}{{\tilde \chi}}
\newcommand{\mlsp}{m_{\tilde \chi}}

\newcommand\one{\leavevmode\hbox{\small1\normalsize\kern-.33em1}}

\newcommand{\met}{\slashchar{p}_T}

\newcommand{\toolfont}[1]{\texttt{#1}}
\newcommand{\mev}{\text{MeV}}
\newcommand{\gev}{\text{GeV}}
\newcommand{\tev}{\text{TeV}}

\newcommand{\iab}{\text{ab}^{-1}}
\newcommand{\ifb}{\text{fb}^{-1}}

\def\slashchar#1{\setbox0=\hbox{$#1$}           
   \dimen0=\wd0                                 
   \setbox1=\hbox{/} \dimen1=\wd1               
   \ifdim\dimen0>\dimen1                        
      \rlap{\hbox to \dimen0{\hfil/\hfil}}      
      #1                                        
   \else                                        
      \rlap{\hbox to \dimen1{\hfil$#1$\hfil}}   
      /                                         
   \fi}

\newcommand{\eg}{\textsl{e.g.}\;}
\newcommand{\ie}{\textsl{i.e.}\;}

\begin{document}

\title{Towards the Final Word on Neutralino Dark Matter}

\author{Joseph Bramante}
\affiliation{Department of Physics, University of Notre Dame, IN, USA}
\author{Nishita Desai}
\affiliation{Institut f\"ur Theoretische Physik, Universit\"at Heidelberg, Germany}
\author{Patrick Fox}
\affiliation{Theoretical Physics Department, Fermilab, Batavia, IL USA}
\author{Adam Martin}
\affiliation{Department of Physics, University of Notre Dame, IN, USA}
\author{Bryan Ostdiek}
\affiliation{Department of Physics, University of Notre Dame, IN, USA}
\affiliation{Department of Physics, University of Oregon, OR, USA}
\author{Tilman Plehn}
\affiliation{Institut f\"ur Theoretische Physik, Universit\"at Heidelberg, Germany}

\begin{abstract}
We present a complete phenomenological prospectus for thermal relic
neutralinos. Including Sommerfeld enhancements to relic abundance and
halo annihilation calculations, we obtain direct, indirect, and
collider discovery prospects for all neutralinos with mass parameters
$M_1,M_2,|\mu| < 4$~TeV, that freeze out to the observed
dark matter abundance, with scalar superpartners decoupled. Much of the relic neutralino sector will be
uncovered by the direct detection experiments Xenon1T and LZ, as well
as indirect detection with CTA.  We emphasize that thermal relic
higgsinos will be found by next-generation direct detection
experiments, so long as $M_{1,2} < 4~{\rm TeV}$. Charged tracks at a 100 TeV hadron collider complement
indirect searches for relic winos.  Thermal relic bino-winos still
evade all planned experiments, including disappearing charged-track
searches. However, they can be discovered by compressed electroweakino
searches at a 100 TeV collider, completing the full coverage of the
relic neutralino surface.
\end{abstract}

\maketitle

\tableofcontents



\section{Introduction}
\label{sec:intro}

While it is sometimes claimed that no physics beyond the Standard Model
need appear below energy scales accessible to imminent experiments, this is not true for
weakly interacting, thermally produced neutralino dark matter.
In the minimal supersymmetric standard model (MSSM), the primary dark matter (DM) 
candidate is the lightest neutralino, which
is an admixture of neutral binos, winos, and higgsinos. Prior studies
have considered which combination of these interaction eigenstates
freeze out to the observed relic abundance. Starting from these pure
states, the relic neutralino surface~\cite{Bramante:2014tba} is limited by abundance criteria
to TeV-scale particle masses, which means that dark matter 
could be unmasked at ongoing direct detection, collider, and indirect
detection experiments~\cite{Baer:1985yd,Barbieri:1988zs,Drees:1992am,Drees:1993bu,Jungman:1995df,Edsjo:1997bg,Ellis:1998kh,Ellis:1999mm,Feng:2000gh,Feng:2000zu,Gounaris:2001fx,Roszkowski:2001sb,BirkedalHansen:2001is,Ellis:2001nx,Baer:2002fv,Ellis:2003ry,Pierce:2004mk,Nihei:2004bc,Baer:2005zc,Baer:2005jq,ArkaniHamed:2006mb,Ellis:2007ac,Berger:2008cq,Hall:2009aj,Feng:2010ef,Cohen:2011ec,Hisano:2011cs,Perelstein:2011tg,Roszkowski:2012uf,Hisano:2012wm,Strege:2012bt,Altmannshofer:2012ks,Grothaus:2012js,Fowlie:2012im,Bechtle:2012zk,Henrot-Versille:2013yma,Han:2013gba,Buchmueller:2013rsa,Boehm:2013qva,Cahill-Rowley:2013vfa,Cohen:2013kna,Chakraborti:2014gea,Cahill-Rowley:2014boa,Roszkowski:2014iqa,Desai:2014uha,Huang:2014xua,Kelso:2014qja,Roszkowski:2014wqa,Buchmueller:2015uqa,Goodsell:2015ura,Hisano:2015rsa}. 

In this work, we advance these phenomenological efforts by including the Sommerfeld
enhancement to thermal freeze-out annihilation for the relic
neutralino surface, \ie $M_1,M_2, |\mu| \lesssim 4~\tev$. This enhancement
substantially alters neutralino masses and experimental prospects
whenever $M_2 \gtrsim 1~\tev$, a region which has often been omitted
in prior work.
In addition, we clarify some facets of relic neutralino phenomenology:
\begin{itemize}
\item It is sometimes stated that future direct detection experiments
  will cover most MSSM neutralino parameter space. We find that a
  preponderance of relic bino-wino parameter space ($M_2 \sim M_1
  \simeq 0.2-2~\tev$ and $|\mu| \gtrsim 2~\tev$) cannot be probed by
  direct, indirect, or LHC searches. The reason is that the lightest supersymmetric 
  partner (LSP) contains only tiny higgsino and wino fractions,
  so its annihilation cross-section, along with
  spin-independent and spin-dependent scattering on nucleons are
  suppressed.
  In addition, the GeV-level bino-wino mass splitting between the lightest chargino (CLSP) and the LSP
  renders collider charged-track searches ineffective.
\item As authors of this paper explored in prior work, the relic
  bino-wino region can be uncovered with compressed electroweakino
  searches at a 100~TeV proton-proton
  collider~\cite{Bramante:2014tba}. We refine these findings for the
  Sommerfeld-enhanced relic neutralino surface in
  Section~\ref{sec:collider}.
\item The well-known systematic uncertainty in the Milky Way's dark
  matter halo density profile obfuscates whether gamma ray searches
  can exclude, have excluded, or will exclude $M_2 \gtrsim 2~\tev$
  thermal relic neutralinos. However, future charged track searches at
  a 100~TeV proton-proton collider will be most sensitive to this
  wino-like LSP parameter space where gamma ray constraints are
  weakest, namely when $|\mu| \sim 4~\tev$ and $M_1 \sim 2-4~\tev$.
\item Contrary to the common lore that higgsino dark matter is
  un-discoverable, we point out that higgsinos that freeze out to the
  observed dark matter relic abundance ($m_{\tilde{H}} \sim 1.1~\tev$)
  will be discovered by next-generation direct detection experiments
  so long as $M_{1,2} \lesssim 4~\tev$.
\end{itemize}
Generally, we find that forthcoming experimental endeavors will be able to probe the
entire relic neutralino surface for $M_1,M_2, |\mu| \lesssim 4~\tev$. Thus it seems that any ``weak scale" MSSM
neutralino sector can be conclusively tested out in the coming
decades~\cite{Cohen:2010gj,Cheung:2012qy,Bramante:2014tba,Badziak:2015qca}.\bigskip

In the remaining sections of this paper, we will explore present and
future experimental probes of MSSM dark matter across the
Sommerfeld-enhanced relic neutralino surface. In each section, we show
how neutralino phenomenology across the surface can be related to
either 
some element of the neutralino and chargino mixing matrices, or a mass
splitting between electroweakino mass eigenstates. Along the way, we
indicate to what extent Sommerfeld-enhanced thermal freeze-out alters
neutralino phenomenology.

In Section~\ref{sec:sommersurface} we introduce the
Sommerfeld-enhanced relic neutralino surface, noting that the
parameter space boundary where the Sommerfeld effect becomes
substantial ($>$TeV mass neutralinos) can be understood as a
consequence of the wino fraction of the LSP, the tree-level wino annihilation
cross-section, and the LSP's freeze-out temperature compared to the mass
of the W and Z bosons \cite{Hisano:2006nn}. In
Section~\ref{sec:direct} we show spin-independent and spin-dependent
direct detection prospects, which are determined by the higgsino and wino
fractions of the LSP, respectively~\cite{Cohen:2010gj}. In
Section~\ref{sec:indirect}, we display the present and future reach of
searches for neutralinos annihilating to gamma rays in the galactic
center, which depends upon the wino fraction of the
LSP. Section~\ref{sec:collider} presents charged track and compressed
$\gamma, \ell, \met$ searches at a 100~TeV collider across the
relic neutralino surface. The charged track search depends on the mass
splitting between the charged lightest supersymmetric partner (CLSP)
and the LSP, while the the mass splitting between the LSP and the next
to lightest neutral supersymmetric partner (NLSP) determines the
efficacy of the compressed search. In Section~\ref{sec:conclusion} we
conclude.

\section{Sommerfelded Relic Neutralino Surface}
\label{sec:sommersurface}

This section introduces the sommerfelded\footnote{From \textit{to   sommerfeld}, \ie enhance through a Sommerfeld factor~\cite{Sommerfeld:1931}. Another possibility would be \textit{sommerfelled} relic neutralino surface, but in spite of the better sound to it we find that this version might be less clear.}
relic neutralino surface and shows that wino-like LSPs will have
enlarged freeze-out annihilation from Sommerfeld-enhancement (SE)\cite{Sommerfeld:1931}.  
Hereafter, we will focus on neutralinos in the MSSM, with all scalar superpartners
decoupled. In our numeric calculations with \toolfont{SuSpect},
\toolfont{microMEGAs}, \toolfont{DarkSUSY}, \toolfont{MG5aMC@NLO}, and
\toolfont{DarkSE} we fix all scalar masses to 8~TeV, including that of
the CP odd Higgs. For 100~TeV proton-proton collider studies, in which
$t$-channel squark exchange with a squark mass of 8~TeV can
substantially increase neutralino production, we remove sfermions entirely.  
For the whole set of neutralino and chargino detection
processes, decoupled squarks present a worst-case scenario, whereas for
specific mixed scenarios, the $s$-channel and $t$-channel contributions
can almost entirely cancel each other. Heavy sfermions are also motivated by models
of split supersymmetry, where most scalar
supersymmetric partners are
decoupled~\cite{Wells:2003tf,ArkaniHamed:2004fb,Giudice:2004tc,Wells:2004di,Kilian:2004uj,Sundrum:2009gv,Arvanitaki:2012ps,Hall:2012zp,Unwin:2012fj,Kahn:2013pfa,Lu:2013cta,Fox:2014moa,Nagata:2014wma,Nomura:2014asa}.\bigskip

Neutralinos in the MSSM are mixtures of the spin-$\frac{1}{2}$
superpartners of the weak gauge bosons, hypercharge gauge bosons, and
Higgs bosons. After electroweak symmetry is broken, the neutral and
charged states mix to form neutralinos and charginos, respectively. We
identify the neutralinos as $\lsp_i^0 = N_{ij}
(\tilde{B},\tilde{W^0},\tilde{H}_u^0,\tilde{H}_d^0)$ and the charginos
as $\lsp_i^{\pm} = V_{ij} (\tilde{W}^\pm, \tilde{H}^\pm)$. Here
$\tilde{B},\tilde{W},\tilde{H}_d^0,\tilde{H}_u^0$, are the bino, wino, and
higgsino fields; $N_{ij}$ and $V_{ij}$ are the neutralino and chargino
mixing matrices in the bino-wino basis, such that $i$ and $j$ index
mass and gauge respectively~\cite{Martin:1997ns}.  The bino, wino, and
higgsino mass parameters are $M_1,~M_2$, and $\mu$, and $\tan \beta$
defines the ratio of up- and down-type Higgs boson vacuum expectation
values in the MSSM.\bigskip

Assuming that all scalar superpartners are heavy, when the universe
cools to $T_\text{rad} < \tev$ during radiation dominated expansion,
MSSM neutralinos freeze out to a relic abundance determined by their
rate of annihilation to Standard Model particles. For neutralinos with
masses below $1~\tev$, it is often sufficient to use tree-level
annihilation cross-sections and ignore the initial state exchange of
photons and weak bosons between annihilating neutralinos. On the other hand, 
the exchange of gauge bosons between two initial-state 
particles can substantially alter the annihilation probability of neutralinos
with masses above $1~\tev$. At threshold this higher-order
correction can diverge like $1/v$, where $v$ is the relative velocity
of the two incoming states. For a Yukawa-like potential, mediated for
example by a $Z$-boson, this effect is cut off at $v \approx
m_Z/\mlsp$, leading to large effects for a large ratio of LSP vs weak
boson masses. This non-relativistic modification of the potential of
two incoming states is called the Sommerfeld effect. 
For freeze-out temperatures below the
mass of electroweak bosons ($T_\text{freeze-out} \equiv \mlsp/20
\lesssim 0.1~\tev$), and thus for lighter LSPs, 
the contribution of $W^\pm$ exchange to the
effective potential of neutralino pairs is suppressed by factors of
$e^{-m_W/T_\text{rad}}$~\cite{Hisano:2006nn}. 

\begin{figure}[t]
\includegraphics[scale=.75]{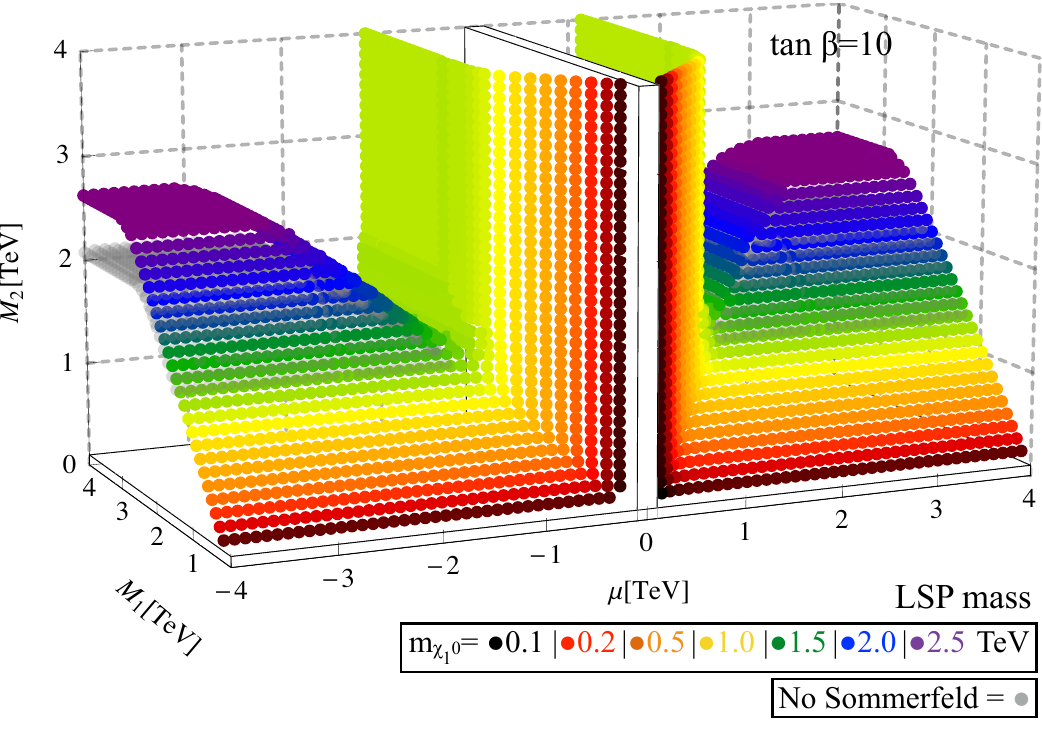}
\includegraphics[scale=.75]{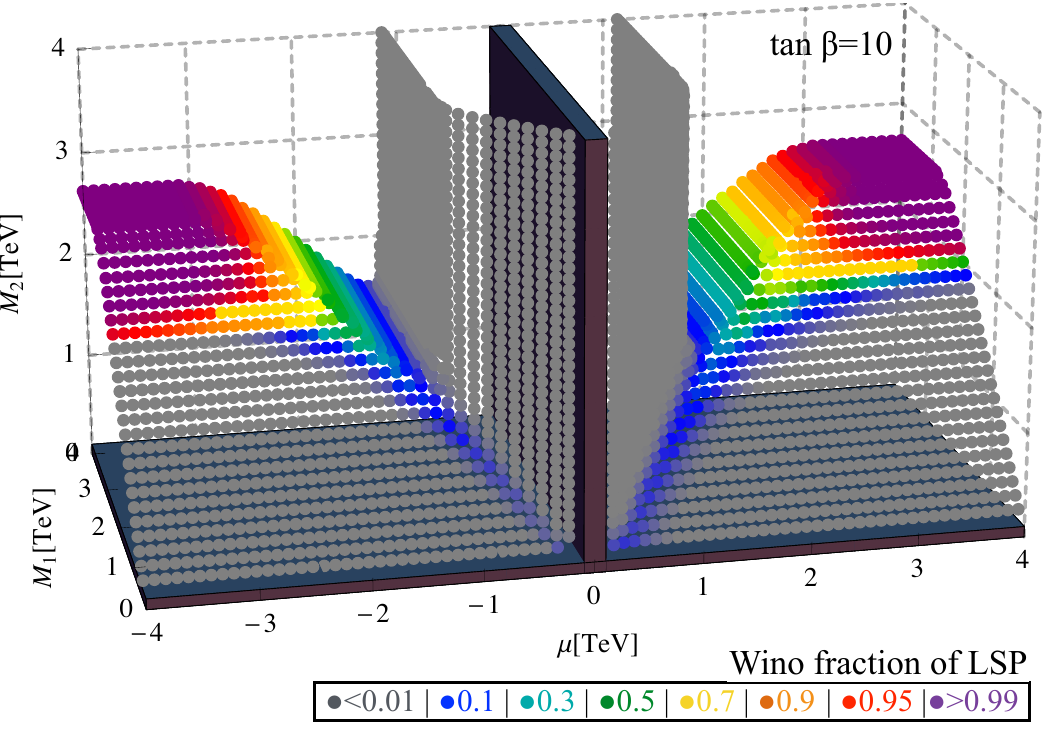}
\caption{\textbf{Left panel:} Combinations of neutralino mass
  parameters $M_1, M_2, \mu$ that produce the correct relic abundance,
  accounting for Sommerfeld-enhancement, along with the LSP mass. The
  relic surface without Sommerfeld enhancement is underlain in
  gray. Regions excluded by LEP are occluded with a white
  box. \textbf{Right panel:} The wino fraction of the lightest
  neutralino.}
\label{fig:lspmass}
\end{figure}

To understand when the Sommerfeld enhancement will affect the
freeze-out of mixed neutralinos, it is useful to first consider the
thermal relic abundance of pure neutralino states. With decoupled
scalars, two neutralinos or charginos can either annihilate through an
$s$-channel $Z$ or Higgs boson, or through a $t$-channel neutralino or
chargino. For the lightest neutralinos the relevant couplings are
\begin{align}
g_{Z \lsp_1^0 \lsp_1^0}
&= \frac{g}{2 \cos \theta_w} \;
   \left( |N_{13}|^2 - |N_{14}|^2
   \right) \notag \\
g_{h \lsp_1^0 \lsp_1^0}
&= \left( g N_{11} - g' N_{12} \right) \; 
   \left( \sin \alpha \; N_{13} + \cos \alpha \; N_{14} \right) \notag \\
g_{W \lsp_1^0 \lsp_1^+}
&= \frac{g \sin \theta_w}{\sqrt{2} \cos \theta_w} \;
   \left( N_{14} V_{12}^* - \sqrt{2} N_{12} V_{11}^*
   \right) \; ,
\label{eq:lsp_couplings}
\end{align}
given in terms of the usual weak gauge couplings, the Higgs mixing angle
$\alpha$, and the neutralino and chargino mixing matrices. 

Obviously pure bino states do not couple to gauge or Higgs bosons, so no
direct annihilation process exists, and their annihilation as well as
Sommerfeld enhancement can only occur through mixing and
co-annihilation.

For pure wino states we need to include the lightest chargino,
typically with a sub-GeV mass difference. Following
Eq.\eqref{eq:lsp_couplings} there will still be no $s$-channel
annihilation process, but for example the LSP can annihilate through
the wino-like chargino in the $t$-channel. Because the two states are
highly mass degenerate, the computation of the current relic abundance
has to include a combined annihilation of the lightest neutralino and
chargino. Neutralino-chargino co-annihilation proceeds through an
$s$-channel $W$ exchange, while diagonal neutralino and chargino
annihilation require a $t$-channel diagram. In the chargino case the
exchange of electroweak bosons between the two non-relativistic incoming particles
leads to a sizeable Sudakov enhancement: an increased
cross section in the numerator of Eq.\eqref{eq:winorelic} has to be
compensated by a larger wino mass on the relic neutralino surface,
\begin{align}
\Omega_{\tilde{W}} h^2 
\simeq 0.12 \left(\frac{\mlsp}{2.1~\tev}\right)^2 
\stackrel{\text{SE}}{\longrightarrow}
0.12 \left(\frac{\mlsp}{2.6~\tev}\right)^2  \; .
\label{eq:winorelic}
\end{align}
In the top panel of Figure~\ref{fig:lspmass} this fact appears
graphically --- the sommerfelded surface, shown with LSP masses
colored, separates from gray points calculated without Sommerfeld enhancement when
$\mlsp \sim 1.5~\tev$, where the wino fraction is sizable.

Finally, pure higgsinos can annihilate efficiently through an
$s$-channel $Z$ diagram. Co-annihilation within the triplet of two
neutralinos and one chargino sets the relic density.
The main distinction between this and the pure wino case, is that chargino pair
annihilation contributes much less to the complete annihilation
process. Because higgsino annihilation is generally more efficient,
and because the contribution of chargino pair annihilation with a
possible electroweak boson exchange between the incoming particles is suppressed,
today's relic density is given by
\begin{align}
\Omega_{\tilde{H}} h^2 
\simeq 0.12 \left( \frac{\mlsp}{1.13~\tev} \right)^2
\stackrel{\text{SE}}{\longrightarrow}
0.12 \left(\frac{\mlsp}{1.14~\tev}\right)^2  \; .
\label{eq:higgsinorelic}
\end{align}
This relatively small effect is hardly visible in
Figure~\ref{fig:lspmass}. There are two reasons why the Sommerfeld enhancement is
significantly larger for the wino case: first, pure chargino
co-annihilation with a photon-induced Sommerfeld effect is roughly
three times more important for pure winos. Second, as previously noted, the $W,Z$-induced
Sommerfeld effect is cut off at $v \approx m_{W,Z}/\mlsp$ 
(compare this to the freeze-out temperature, $\sim \mlsp/20$), which means
that it influences more phase-space for pure winos at freeze-out.\bigskip

To generate the sommerfelded surface shown in Figure~\ref{fig:lspmass}, we
first calculate electroweakino mass parameters with
\toolfont{SuSpect}~\cite{Djouadi:2002ze}. We include the loop-level,
custodial-symmetry-breaking-induced mass separation between the
charged and neutral components of both the wino and higgsino, setting
these to 160~MeV~\cite{Feng:1999fu,Gherghetta:1999sw,Ibe:2012sx} and
350~MeV~\cite{Cheng:1998hc,Fritzsche:2002bi,Cirelli:2005uq}
respectively, before diagonalizing electroweakino mass matrices. As we
discuss further in Section~\ref{sec:collider}, the values of
electroweakino mass parameters can also substantially shift these
charged-neutral mass splittings. With this electroweakino mass
spectrum, we require each point to satisfy $\Omega_\lsp h^2 \simeq
0.12 \pm 0.005$, calculating the sommerfelded relic abundance
using \toolfont{DarkSE}~\cite{Hryczuk:2011vi,Hryczuk:2014hpa}, which
improves upon the relic density calculations of
\toolfont{DarkSUSY}~\cite{Bechtle:2012zk}, and includes Sommerfeld contributions
to each LSP annihilation channel, for up to three charge-equivalent initial state pairs
of electroweakinos.

As a comparison to the relic neutralino surface in
Ref.~\cite{Bramante:2014tba}, we also calculate the
sommerfelded surface in the pure wino approximation using
\toolfont{microMEGAs} and following the procedure in
Ref.~\cite{Beneke:2014gja}. Without Sommerfeld enhancement, the
calculated relic density differs between the two programs by about
10\%, with \toolfont{microMEGAs} giving the higher number.  After
including the Sommerfeld enhancement, the maximal wino-like LSP mass
from \toolfont{microMEGAs} is 2.5~TeV, compared to 2.6~TeV from
\toolfont{DarkSE}.

\section{Direct Detection}
\label{sec:direct}

\begin{figure}[t]
\includegraphics[scale=.75]{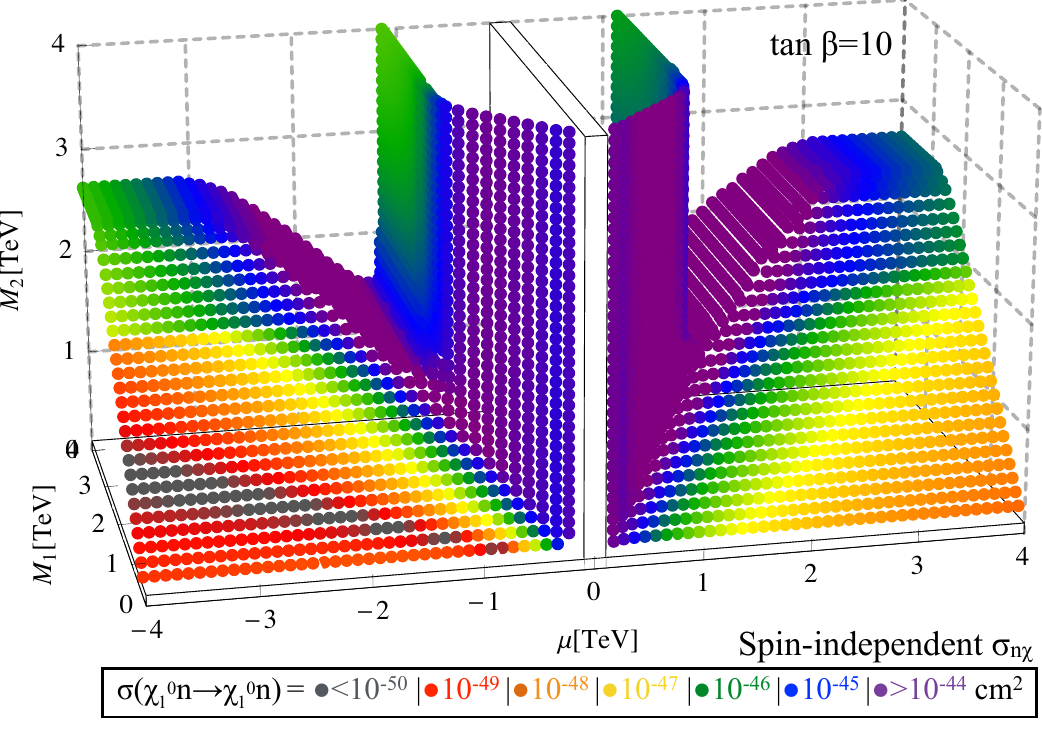}
\includegraphics[scale=.75]{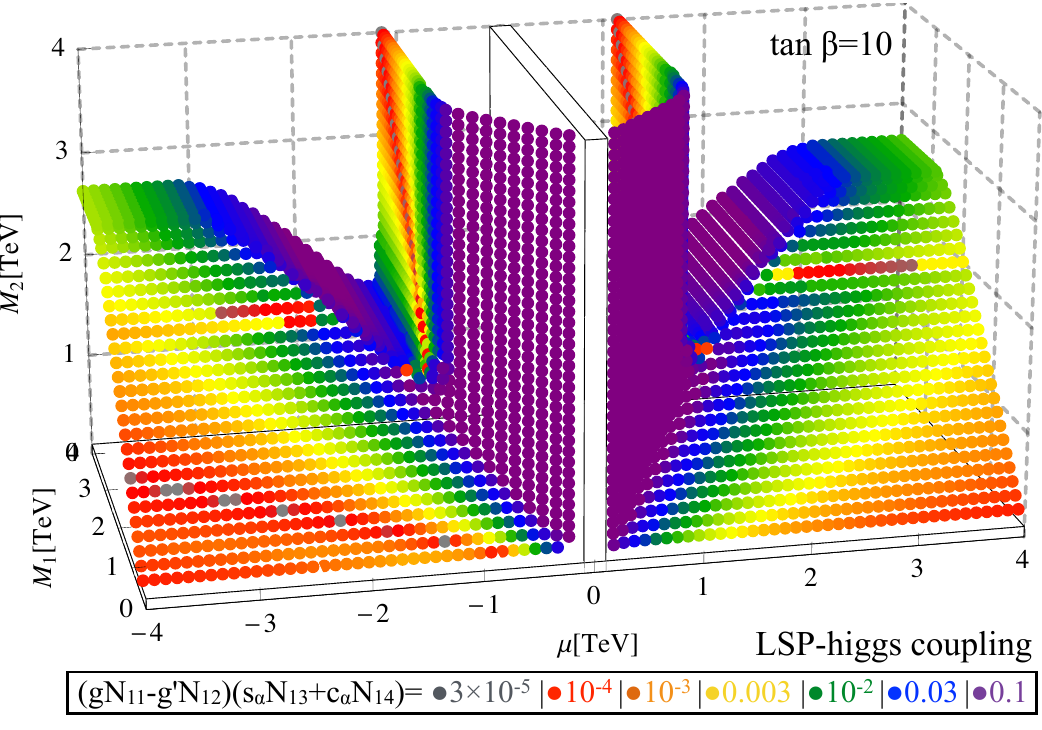} \\
\includegraphics[scale=.75]{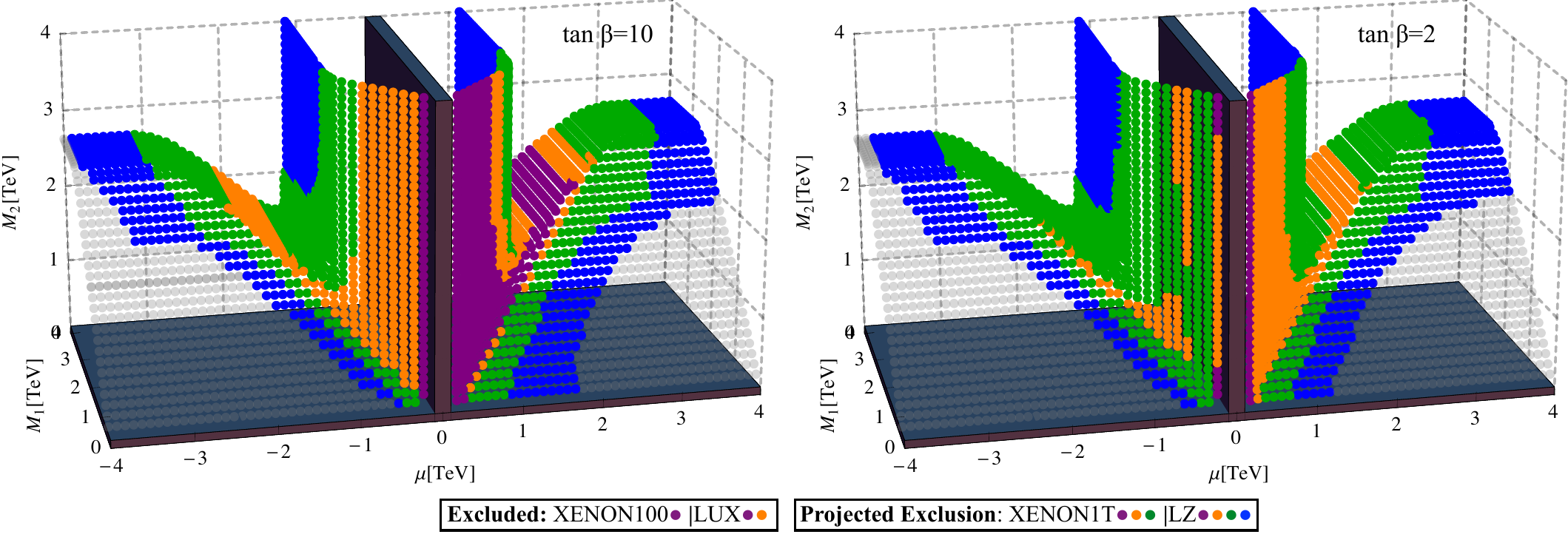}
\caption{\textbf{Top left panel:} The spin-independent
  nucleon-scattering cross-section for relic neutralinos is shown, as
  calculated by
  \toolfont{microMEGAs}~\cite{Belanger:2013oya}. \textbf{Top right
    panel:} The coupling of neutralinos to the SM-like
  MSSM Higgs boson. \textbf{Bottom left panel:} Relic neutralino
  exclusions from XENON100 and LUX and prospects from XENON1T and LZ
  for tan $\beta=10$. The boxed out area denotes the LEP
  exclusion. \textbf{Bottom right panel:} The same for tan $\beta=2$.}
\label{fig:spininde}
\end{figure}

Detection of neutralinos via nuclear scattering experiments can be
divided into two categories: spin-independent (SI) and spin-dependent
(SD). In the spin-independent case, neutralinos will scatter off
nucleons via the exchange of a Higgs boson, which couples to quarks and
quark loops within nucleons in atomic
nuclei~\cite{Hisano:2010ct,Hisano:2011cs,Hill:2013hoa,Hill:2014yka,Hill:2014yxa,Hisano:2015rsa}. Following
Eq.\eqref{eq:lsp_couplings} this coupling is driven by bino--higgsino
and wino--higgsino mixing.
The spin-independent scattering will be maximized when the LSP is an
even bino--higgsino or wino--higgsino mixture.  Providing
confirmation, Figure~\ref{fig:spininde} shows that the LSP-Higgs
coupling is indeed proportional to the size of SI neutralino-nucleon
scattering over the relic neutralino surface --- and that SI scattering
cross-sections reach their apex on the bino-higgsino and wino-higgsino
slopes, $M_{1,2} \sim \mu$. The apparent proportionality between
coupling and scattering would be more exact if we incorporated the
small but non-negligible contribution of the heavy Higgs
bosons. 

The SI neutralino-nucleon cross-sections in
Figure~\ref{fig:spininde} are obtained from
\toolfont{microMEGAs}~\cite{Belanger:2013oya}.  The lower panels of
Figure~\ref{fig:spininde} also display the current exclusions on
spin-independent neutralino-nucleon scattering from
Xenon100~\cite{Aprile:2012nq} and LUX~\cite{Akerib:2013tjd}, along
with projected exclusions from Xenon1T and
LZ~\cite{Cushman:2013zza}. These projections indicate that all relic
neutralinos lighter than 4~TeV, except a large swathe of bino-winos
(addressed in Section~\ref{sec:collider}), will be probed by upcoming
direct detection experiments. 

The cross-sections found for SI scattering here match to within a factor of two, recent studies of
neutralino-nucleon scattering in a particular decoupling limit, Refs.~\cite{Hill:2013hoa,Hill:2014yka,Hill:2014yxa}.\footnote{
In our work, the least coupled higgsino-like LSP point shown in
Figure \ref{fig:spininde} has a cross-section,
\begin{align*}
(M_1 =4~{\rm TeV}, M_2 =4~{\rm TeV}, |\mu| =1.1~{\rm TeV}) \rightarrow \sigma_{n\tilde{\chi}}^{\rm (SI)} \simeq 10^{-46}~{\rm cm^2}.
\end{align*}
For this point, matching Eq.~\eqref{eq:lsp_couplings} to the higgs-LSP coupling of Ref.~\cite{Hill:2013hoa}, and using this to determine $\kappa$ in Ref.~\cite{Hill:2013hoa} yields
\begin{align*}
(M_1 \rightarrow \infty, M_2 =4~{\rm TeV}, |\mu| =1.1~{\rm TeV}) &\rightarrow \sigma_{n\tilde{\chi}}^{\rm (SI)}\simeq 7 \times 10^{-47}~{\rm cm^2} \\
(M_1 =4~{\rm TeV}, M_2 \rightarrow \infty, |\mu| =1.1~{\rm TeV}) &\rightarrow \sigma_{n\tilde{\chi}}^{\rm (SI)}\simeq 7 \times 10^{-47}~{\rm cm^2}.
\end{align*}}
Even taking $M_1\rightarrow \infty$ or $M_2\rightarrow \infty$ as in \cite{Hill:2013hoa}, the resulting $1.1$ TeV mass higgsino appears to be within reach of LZ \cite{Cushman:2013zza}, so long as $M_2<4 {~\rm TeV}$ or $M_1<4 {~\rm TeV}$.\bigskip

\begin{figure}[t]
\includegraphics[scale=.75]{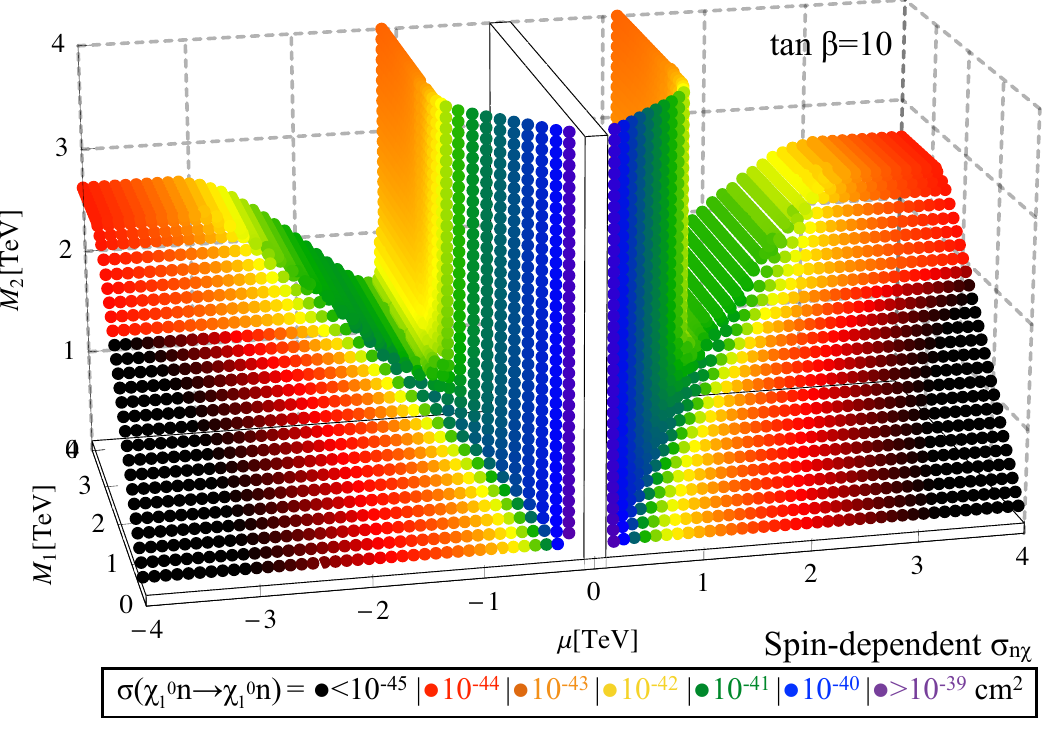}
\includegraphics[scale=.75]{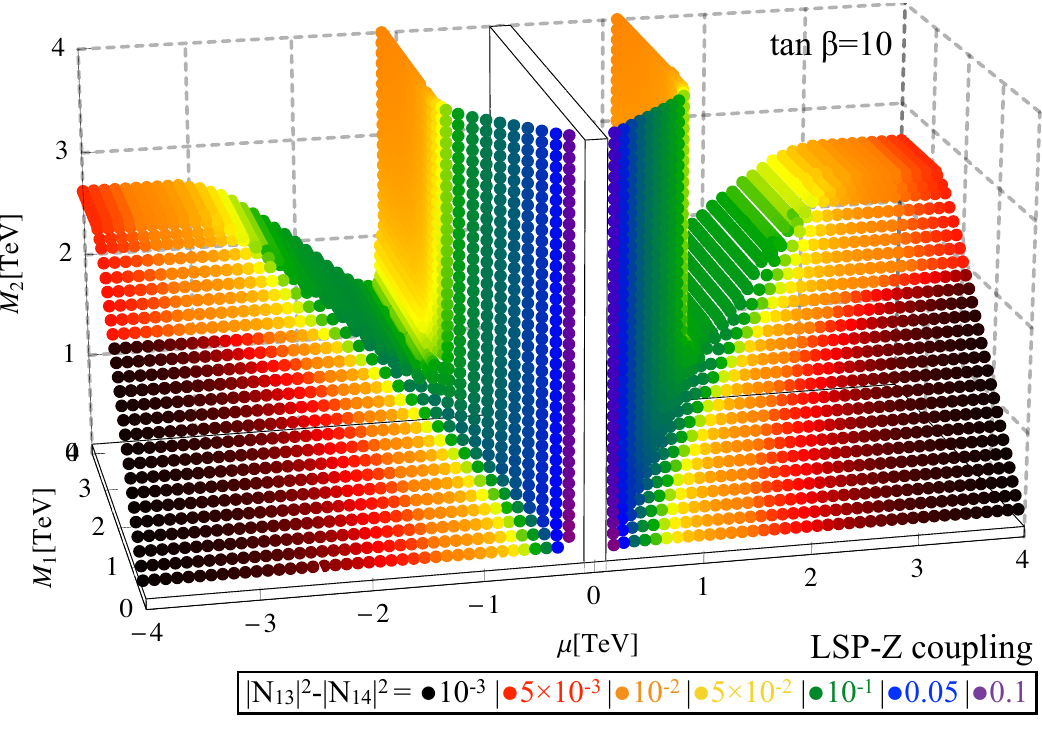}
\caption{\textbf{Left panel:} The spin-dependent nucleon-scattering
  cross-section for relic neutralinos, as calculated by
  \toolfont{microMEGAs}~\cite{Belanger:2013oya}. \textbf{Right panel:}
  The proportional coupling of neutralinos to the $Z$ boson.}
\label{fig:spinde}
\end{figure}

In the case of spin-dependent scattering, which occurs through
$Z$-boson exchange, and thus depends upon the spin of the nuclear
scattering target, the detection of neutralinos depends solely on the
higgsino fractions of the neutralino (\ie what portions are $H_u,
H_d$). As shown in Eq.\eqref{eq:lsp_couplings}, binos and winos do not
couple to the $Z$ boson. Moreover, if $|N_{13}|=|N_{14}|$, which happens for pure Higgsinos,
the neutralino spin-dependent scattering cross-section vanishes. In
Figure~\ref{fig:spinde} we show the spin-dependent neutralino-nucleon
scattering cross-section, as well as the LSP-$Z$ coupling across the
relic neutralino surface. The correspondence is striking --- the size
of the $Z$-neutralino coupling determines the size of the
spin-dependent cross-section. Constraints from the current generation
of spin-dependent scattering of relic neutralino dark matter at
experiments like SIMPLE~\cite{Felizardo:2011uw},
COUPP~\cite{Behnke:2012ys}, Xenon100~\cite{Aprile:2013doa}, and
PICO2L~\cite{Amole:2015lsj}, are less stringent than spin-independent
constraints. However, future experiments like
PICO250~\cite{Cushman:2013zza} will be able to probe TeV-mass thermal
relic bino-higgsinos.

\section{Indirect Detection}
\label{sec:indirect}

Gamma ray surveys of the galactic center have bounded dark matter
annihilation to photons, $\lsp_1^0 \lsp_1^0 \rightarrow \gamma \gamma,
Z \gamma$, or intermediate particles which decay to photons, $\lsp_1^0
\lsp_1^0 \rightarrow W^+ W^-$. However, these bounds remain somewhat
uncertain, because they depend upon the Milky Way's DM density
profile.  The flux of photons $\Phi_\gamma$ arising from dark matter
annihilating inside an observed cone of solid angle $\Delta \Omega$ is
\begin{align}
\frac{d \Phi_\gamma}{d E_\gamma} 
= \frac{\left\langle \sigma v \right\rangle}{8 \pi m_X^2} \;
  \frac{d N_\gamma}{d E_\gamma} \; 
  \int_{\Delta \Omega} d \Omega \int_\text{line of sight} d l ~\rho_\lsp^2(l) \; ,
\label{eq:photflux}
\end{align}
where $E_\gamma$ is the energy of the photon, $\left\langle \sigma v
\right\rangle$ is the averaged DM annihilation cross-section,
$N_\gamma$ is the number of photons produced per annihilation, and
$l$ is the distance from the observer to the DM annihilation event.

\begin{figure}[b!]
\includegraphics[scale=.75]{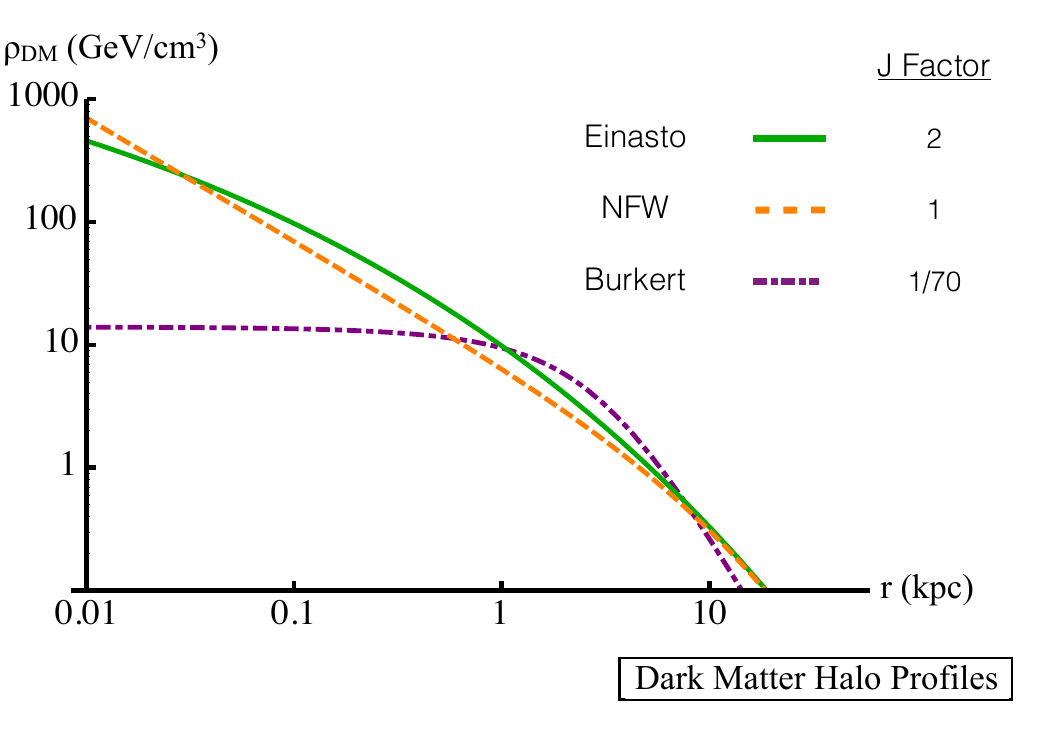}
\caption{Dark matter galactic halo profiles, including standard
  Einasto and NFW profiles along with a Burkert profile with a 3 kpc
  core. $J$ factors are obtained assuming a spherical DM distribution
  and integrating over the radius from the galactic center from
  $r\simeq 0.05$ to 0.15~kpc. $J$ factors are normalized so that
  $J(\rho_\text{NFW})=1$.}
\label{fig:halo_profiles}
\end{figure}

Because Eq.\eqref{eq:photflux} is proportional to $\rho_\lsp^2$,
any annihilation constraint relies on assumptions about the Milky Way's
DM density profile. Assuming a steeper DM halo profile, \ie DM density
increasing more rapidly towards the core of the Milky Way, results in
a more stringent bound on DM annihilation.  We consider three DM halo
density profiles that are increasingly flat towards the center of the
Milky Way. The generalized NFW profile~\cite{Navarro:1995iw} is
given by
\begin{align}
\rho_\text{NFW}(r) = \frac{\rho_\odot}{\left(r/R\right) \left(1+r/R\right)^2},
\end{align}
where $r$ is the distance from the galactic center, and we assume a
characteristic scale $R=20~\text{kpc}$, solar position DM density
$\rho(r_{\odot}) \equiv 0.4~\gev/\text{cm}^3$, and $r_\odot
=8.5~\text{kpc}$ throughout this study. Second, we consider the
Einasto profile,
\begin{align}
\rho_\text{Ein}(r) = \rho_\odot~ \exp \left[-\frac{2}{\alpha} \left(\left( \frac{r}{R}\right)^\alpha -1\right) \right],
\end{align}
where we take $\alpha=0.17$ and $R=20~\text{kpc}$. This is the halo
profile model that best fits micro-lensing and star velocity
data~\cite{Iocco:2011jz,Pato:2015dua}.  Third, we consider a Burkert
or ``cored" profile, with constant DM density inside radius
$r_c=3~\text{kpc}$,
\begin{align}
\rho_\text{Burk}(r) =
\frac{\rho_\odot}{\left(1+r/r_c\right) \left(1+(r/r_c)^2\right)},
\end{align}
For this profile, $r_c$ sets the size of the core --- we assume
$r_c=3~\text{kpc}$. Assuming such a large core results in very diffuse
dark matter at the galactic center, and therefore yields the weakest
bound on neutralino self annihilation. On the other hand, assuming a
core of smaller size (\eg 0.1~kpc) only alters DM annihilation
constraints by an $\mathcal{O}(1)$ factor~\cite{Hooper:2012sr}. 

In Figure~\ref{fig:halo_profiles}, we illustrate the three halo profiles.
The impact on gamma ray flux of different dark matter halo profiles is
conveniently parameterized with a $J$ factor,
\begin{align}
J \propto \int_{\Delta \Omega} d \Omega \int_{l.o.s.} d {l} ~\rho_\lsp^2(l) \sim \int d r ~\rho_\lsp^2(r).
\end{align}
We show $J$ factors integrating over the approximate H.E.S.S. galactic
center gamma ray search range, $r\simeq0.05$ to 0.15~kpc, and
normalizing so that $J(\rho_\text{NFW})=1$.\bigskip

\begin{figure}[t]
\includegraphics[scale=.75]{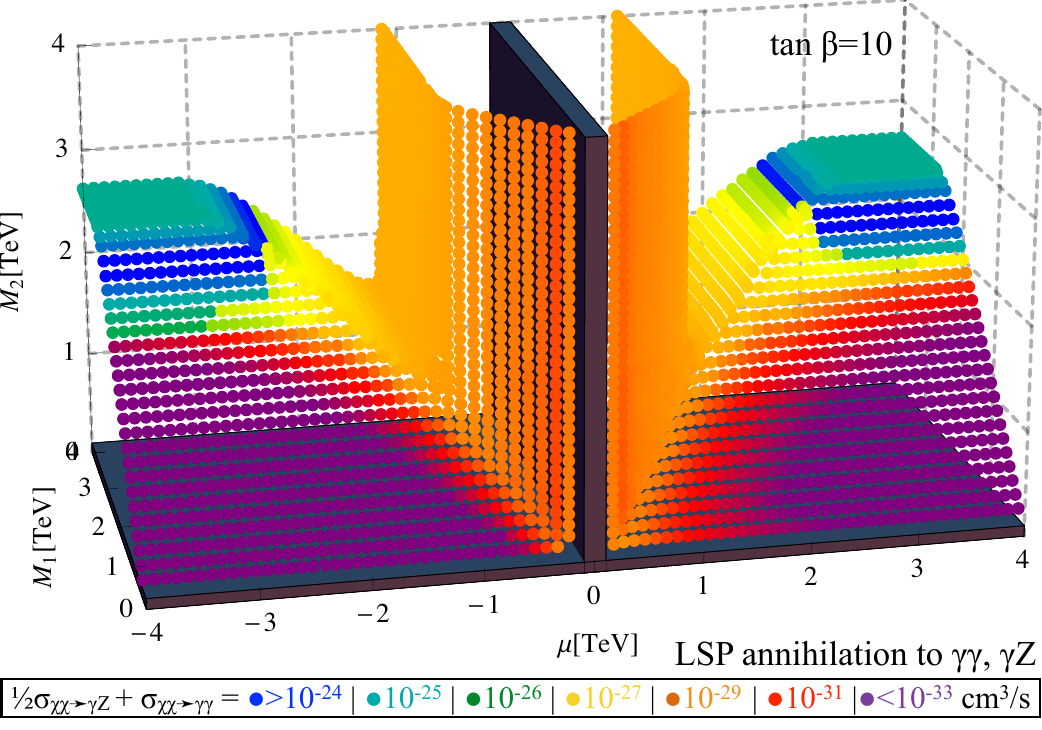}
\includegraphics[scale=.75]{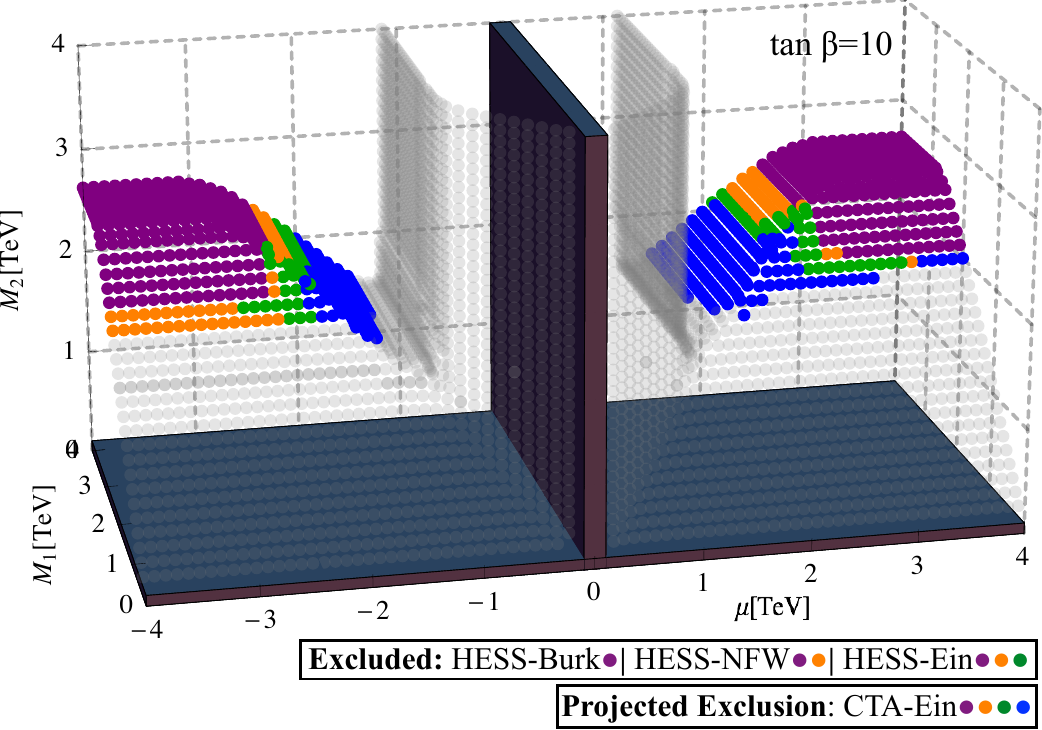}
\caption{\textbf{Left panel:} The neutralino annihilation
  cross-section to $\gamma \gamma$ and $\frac{1}{2}$Z$\gamma$ is given
  for Milky Way velocities, as detailed in the text. \textbf{Right
    panel:} Relic neutralino parameters excluded by the HESS gamma ray
  line search, assuming Einasto, NFW, and cored (Burkert, 3 kpc)
  profiles, along with the projected CTA exclusion for an Einasto
  profile.}
\label{fig:annihilation}
\end{figure}

Galactic center gamma ray bounds on MSSM neutralinos depend on our
knowledge of the cross-sections for neutralino annihilation to
electroweak bosons.  Neutralino annihilation rates to photons and Z
bosons are known including one-loop
corrections~\cite{Bergstrom:1997fh,Bern:1997ng,Ullio:1997ke,Boudjema:2005hb}.
In addition, neutralinos annihilating non-relativistically with masses
greater than $\sim~\tev$ will again exhibit a Sommerfeld
enhancement~\cite{Hisano:2002fk,Hisano:2003ec,Hisano:2004ds,Hisano:2006nn,Slatyer:2009vg,Hryczuk:2010zi,Hryczuk:2011tq,Beneke:2012tg,Hellmann:2013jxa,Beneke:2014gja,Beneke:2014hja}. This
can enhance pure wino annihilation to photons and weak bosons by
orders of magnitude for $\mlsp = 1-5~\tev$ with a typical Milky Way DM
velocity $v \sim
0.001$~\cite{Ovanesyan:2014fwa,Baumgart:2014vma,Bauer:2014ula}.\bigskip

While a number of papers have addressed galactic center constraints
including sommerfelded pure
winos~\cite{Cohen:2013ama,Fan:2013faa,Baumgart:2014saa}), we provide
indirect bounds on mixed neutralinos. We use the following
prescription: if the neutralino LSP is more than $90\%$ wino
($N_{12}^2 > 0.9$), we use the sommerfelded, pure wino one-loop
results of Ref.~\cite{Hryczuk:2011tq} for $\sigma_{\lsp \lsp
  \rightarrow \gamma \gamma}$, $\sigma_{\lsp \lsp \rightarrow \gamma
  Z}$, and $\sigma_{\lsp \lsp \rightarrow W^+ W^-}$. If the neutralino
LSP is less than $90\%$ wino we compute these cross-sections with
\toolfont{micrOMEGAs4}~\cite{Belanger:2014vza}, which utilizes
one-loop
results~\cite{Bergstrom:1997fh,Bern:1997ng,Ullio:1997ke,Boudjema:2005hb}. Because
\toolfont{micrOMEGAs4} does not include Sommerfeld enhancement for
neutralino parameter space where ($N_{12}^2 < 0.9$), this prescription
produces conservative bounds.

In Figure~\ref{fig:annihilation} we indicate bounds on relic
neutralino dark matter from gamma ray line searches conducted by
H.E.S.S.~\cite{Abramowski:2013ax} along with those projected for the
Cerenkov Telescope Array (CTA)~\cite{Bergstrom:2012vd} (see also
HAWC~\cite{Abeysekara:2014ffg}). We vary the dark matter
profiles. Excluding pure wino dark using HESS and Fermi-LAT
data, assuming Einasto or NFW profiles, has been studied extensively, 
in e.g. \cite{Cohen:2013ama,Fan:2013faa,Baumgart:2014saa}. The
right panel of Figure~\ref{fig:annihilation} shows that, for mixed
electroweakinos, wino-like LSPs with a small bino or higgsinos component and mass
above $2~\tev$ can be excluded under the assumption of an Einasto or
NFW profile. However, the assumption of a more cored profile lifts
bounds on some heavier relic bino-winos and wino-higgsinos. Comparing
the LSP wino fraction in Figure~\ref{fig:lspmass} with
Figure~\ref{fig:annihilation} shows that exclusions on relic
neutralino annihilation increase with wino fraction. It is also
interesting to note that, under the assumption of an Einasto
profile~\cite{Iocco:2011jz,Pato:2015dua}, CTA will probe the entire
wino-higgsino surface, and all bino-winos for which the LSP is
wino-like.

\section{100~TeV Collider}
\label{sec:collider}

\begin{figure}[t]
\includegraphics[scale=.75]{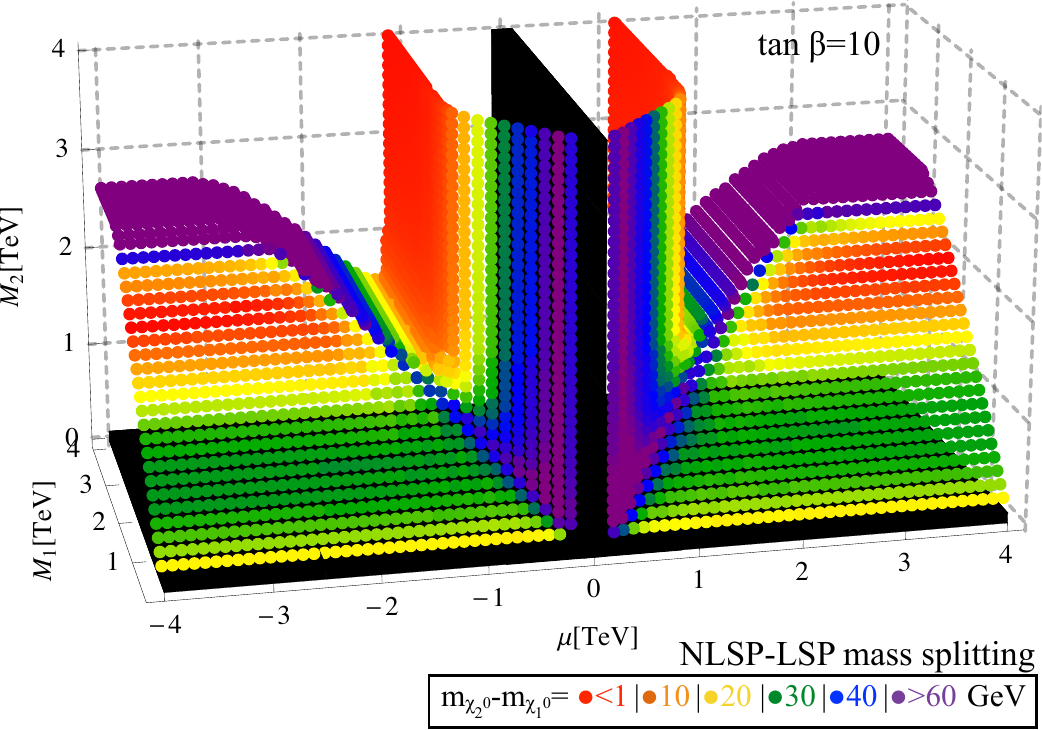}
\includegraphics[scale=.75]{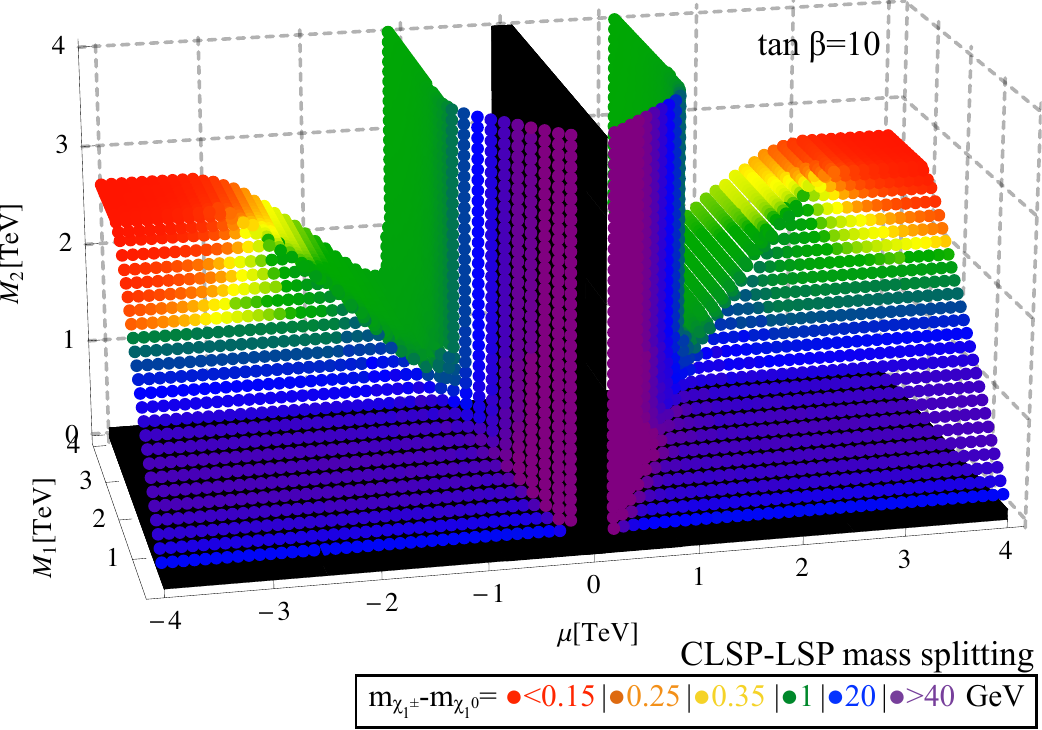}
\caption{\textbf{Top panel:} The mass splitting between the
  NLSP and LSP. \textbf{Right
    panel:} The mass splitting between the lightest chargino (CLSP)
  and lightest neutralino (LSP). Parameters excluded by LEP are
  occluded with a black box. If the CLSP-LSP mass splitting is below
  roughly 1~GeV, the point is accessible with charged track searches;
  if CLSP-LSP and NLSP-LSP mass splittings are between 10-60~GeV,
  the point is accessible with compressed electroweakino searches.}
\label{fig:msplittings}
\end{figure}

Because most models of dark matter are weakly-coupled to Standard
Model particles, generic collider dark matter searches focus on events
with large missing transverse momentum ($\met$), arising when weakly-interacting dark matter
recoils off Standard Model particles (i.e. jets, photons, leptons). On the other hand,
collider searches directed at a relic, co-annihilating neutralino--chargino sector
benefit from searching for electroweak radiation, emitted in
inter-electroweakino decays.

For a nearly pure wino LSP, almost mass-degenerate charginos decaying to
neutralinos deposit electroweak radiation as \textsl{charged
  tracks}. Around the wino plateau, 
  the mass splitting between the lightest chargino and the lightest neutralino becomes compressed, as
shown in Figure~\ref{fig:msplittings}. For these points, the
chargino-neutralino mass difference is set by loop effects, the
chargino-neutralino decay width decreases, and the chargino lifetime
is long enough for the chargino to leave noticeable paths in the
detectors. Thus, typical mass splittings around $100~\mev$ shown in
Figure~\ref{fig:msplittings} are ideal for disappearing charged track
searches~\cite{Feng:1999fu, Gunion:1999jr, Abreu:1999qr,
  Acciarri:2000wy, Abbiendi:2002vz, Ibe:2006de, Asai:2008sk,
  Buckley:2009kv, Aad:2013yna, Low:2014cba,Cirelli:2014dsa,
  CMS:2014gxa,Ostdiek:2015aga}.

Recently, a number of strategies for \textsl{compressed electroweakino
  searches} have been developed, targeting supersymmetric dark matter
with $10-60~\gev$ inter-electroweakino mass
splittings~\cite{Schwaller:2013baa,Han:2014kaa,Bramante:2014dza,Han:2014xoa,Low:2014cba,Baer:2014kya,Bramante:2014tba,Han:2015lma,Han:2015lha,Bhattacharyya:2009cc,Giudice:2010wb,Calibbi:2013poa,Gori:2014oua,Chakraborti:2015mra}. These
searches require a half to a fifth of the $\met$ required by
straightforward jet plus $\met$ searches, but add the requirement of
$p_T \sim 10-60~\gev$ photons and leptons, which appear in the
electroweakino decays.\bigskip

The small mass splittings between electroweakinos, utilized by
compressed and charged track searches, are a consequence of requiring
that they freeze-out to the observed dark matter relic abundance with
the help of co-annihilation processes. For co-annihilation to
contribute significantly to the LSP annihilation, the CLSP or NLSP state must
be abundant in the thermal bath when the LSP freezes out --- so smaller NLSP-LSP
and CLSP-LSP mass splittings increase co-annihilation. Partly because of this, nearly-pure winos, with
a chargino-neutralino mass splitting of 160 MeV, are
the most massive thermal relic neutralinos. In the case of
bino-wino neutralinos with $M_2 < 2~\tev$, where the LSP is bino-like, the NLSP-LSP masses cannot be
further apart than $m_{\lsp_2^0}-m_{\lsp_1^0}=10-40~\gev$. Figure~\ref{fig:msplittings}
illustrates this point, showing that precisely the regions
inaccessible to direct (Figure~\ref{fig:spininde}), indirect
(Figure~\ref{fig:annihilation}), and present collider searches, could
be tested by compressed electroweakino searches \cite{Bramante:2014tba} at a 100~TeV
proton-proton collider~ \cite{Yu:2013wta,Anderson:2013ida,Cohen:2013xda,Curtin:2014jma,Bramante:2014uda,Fowlie:2014awa,Rizzo:2014xma,Larkoski:2014bia,Hook:2014rka,Barr:2014sga,Cohen:2014hxa,Acharya:2014pua,Curtin:2014cca,diCortona:2014yua,Barr:2014sga,Berlin:2015aba,Kotwal:2015rba,Plehn:2015cta,Harris:2015kda,Hajer:2015gka,Ellis:2015xba}.

\subsection{Charged track search}

The disappearing charged track search strategy relies on an enhanced
lifetime of charginos which are around $100~\mev$ heavier than the
dark matter agent. When the mass difference is below $1~\gev$, the
dominant decay mode is $\lsp^\pm_1 \rightarrow \lsp^0_1 \pi^\pm$,
which is not calculated in many of the publicly available SUSY decay
codes. Using the procedure detailed in Section~\ref{sec:sommersurface}
we determine which points on the relic neutralino surface have a mass
splitting smaller than 1~GeV and calculate their chargino-neutralino
decay widths based on Ref.~\cite{Chen:1996ap}. The resulting decay
lengths range from 1-50 mm, for these points on the relic neutralino surface.
Thus, even before a possible boost is taken into account, many
of these charginos travel macroscopic
distances before decaying. The neutralino takes the majority of the
momentum of the decay products, leaving the pion with too little
energy to be seen. The result is a charged track which disappears
without leaving deposits of energy in the calorimeters.\bigskip

To begin our study, we first calibrate our method based on the ATLAS
search for disappearing tracks at 8~TeV~\cite{Aad:2013yna}. We study a
simplified model in which the chargino is 160~MeV heavier than the
neutralino and has a lifetime of 0.2~ns. We generate all combinations
of chargino production with up to two extra partons in the final state
using \toolfont{MG5aMC@NLO}~\cite{Alwall:2014hca}. These events are
then showered, matched, and hadronized using
\toolfont{Pythia6.4}~\cite{Sjostrand:2006za} with the MLM matching
scheme~\cite{Mangano:2006rw}. Finally, they are passed through
\toolfont{DELPHES3}~\cite{deFavereau:2013fsa} using the default ATLAS
card.  Jets are defined using the anti-$k_T$
algorithm~\cite{Cacciari:2008gp} with $R= 0.4$ as implemented in
\toolfont{FastJet}~\cite{Cacciari:2011ma} and are then required to
have $p_{T,j} > 20~\gev$ and $|\eta_j| < 2.8$. The signal also
requires a lepton veto; electron candidates are defined with $p_{T,e}
> 10~\gev$ and $|\eta_e| < 2.47$ while muon candidates are also
defined with $p_{T,\mu} > 10~\gev$ but $|\eta_\mu| < 2.4$. Following the ATLAS
jet and lepton definition protocol \cite{Aad:2013yna}, 
to enforce lepton isolation we remove any jet candidate
within $\Delta R_{j\ell}<0.2$ of a lepton, from jet candidates. 
After this, any lepton within $\Delta R=0.4$ of remaining jet 
candidates is incorporated into that jet.

The final object needed for the search is the disappearing
track. While \toolfont{DELPHES} details charged final states with
an $\eta$ and $p_T$ dependent efficiency, the charginos are not
considered a final state. \toolfont{Pythia} does propagate the chargino, but it
does not include the effect of the magnetic field. This should have
little effect as the charginos are typically boosted enough that their tracks
can be reconstructed \cite{Aad:2013yna}. As such, we take the final location of the
chargino and compute the transverse length traveled.  To count as an
isolated track, there must also be no jets with $p_{T,j} > 45~\gev$
within $\Delta R_{j \, \tr} =0.4$.  Moreover, the sum of the $p_T$ of
all tracks with $p_T > 400~\mev$ and within a cone of $\Delta R=0.4$
is required to be less than $4\%$ of the $p_T$ of the candidate
isolated track. Finally, the considered chargino track must have the
highest $p_T$ of all isolated tracks.\bigskip

To extract the signal ATLAS then applies a series of cuts:
\begin{enumerate}
	\item leading jet $p_{T,j}>90~\gev$
	\item missing transverse momentum $\met > 90~\gev$
	\item $\Delta \phi_{j,\met} > 1.5$. For extra jets with
          $p_{T,j} > 45~\gev$ this applies to the leading two jets.
	\item isolated track with transverse length $=30-80$~cm
        \item $p_{T,\tr} > 15~\gev$ and $0.1 < |\eta_\tr| < 1.9$.
\end{enumerate}
Before applying the last cut, ATLAS provides a benchmark for a 200~GeV
chargino: with $20.3~\ifb$ of integrated luminosity, they obtain 18.4
Monte Carlo events passing all other cuts. In our simulation, 23.9
events pass. We take the corresponding ratio $\epsilon_\tr = 0.77$ as a
flat efficiency for measuring a disappearing track with $0.1 < |\eta|
< 1.9$ and $p_T> 15~\gev$, and a track length between 30 and 80 cm.
The visible cross section is then defined as
\begin{align}
\sigma_\text{vis} = \sigma_\text{MC} \times \epsilon_\text{cuts} \times \epsilon_\tr \; .
\label{eqn:crosssection}
\end{align}
The background for a disappearing track search is complex, 
because it is not dominated by a Standard Model
process. Instead, it is very detector dependent and
involves charged hadrons interacting with the detector material with
large momentum exchange and $p_T$-mis-measured tracks. 
ATLAS makes a measurement
of the $p_T$-mis-measured tracks and fits the shape as
$d\sigma/d p_{T,\tr} = 1/p_{T,\tr}$ where
$a=1.78\pm0.05$. Following the example of Refs.~\cite{Low:2014cba}
and~\cite{Cirelli:2014dsa}, we normalize this to the total background
of 18 events with $p_{T,\tr} > 200~\gev$ with $20.3~\ifb$
of data.\bigskip

Extrapolating this search to a 100~TeV collider requires some
assumptions. First, since the background is detector dependent, we
conservatively choose a default ATLAS setup and detector card in
\toolfont{DELPHES}.

We assume that the efficiency for detecting these disappearing tracks
remains at a constant $\epsilon_\tr = 0.77$ across the range of
parameters. Furthermore, we assume that the shape of the background
remains the same at 100~TeV collisions as it was at 8~TeV. This
assumption can be tested at the 13~TeV run of the LHC.  The background
normalization we use rescales the background found at ATLAS,
by using the ratio of the $Z(\nu\bar{\nu})$+jets cross sections
that pass initial analysis cuts on $p_{T,j}$, $\met$, and $\Delta
\phi_{j,\met}$, at $\sqrt{s} =$ 8~TeV and 100~TeV, respectively.

The same steps are used in Refs.~\cite{Low:2014cba}
and~\cite{Cirelli:2014dsa} to estimate the background for the
disappearing track signature at a 100~TeV collider. Both references
acknowledge the large amount of uncertainty and present their searches
for the pure wino as a band with the background $20\%$ to $500\%$ as
large as the estimated value. Both find that a pure wino could be
discovered at the 100~TeV collider, although
Ref.~\cite{Cirelli:2014dsa} uses different cuts, resulting in improved
discovery prospects. Here we combine these searches with the
constrains from the observed dark matter relic abundance, including
slightly mixed binos.  To this end, we use the optimized cuts of
Ref.~\cite{Cirelli:2014dsa} and scan over a representative sample of
the relic neutralino surface. The optimized cuts are
\begin{align}
	p_{T,j_1} &> 1~\tev  &p_{T,j_2} > 500~\gev \notag \\
	\met &> 1.4~\tev  &p_{T,\tr} > 2.1~\tev \; ,
\end{align}
All other cuts are identical to the ATLAS analysis. For each of the
data points we calculate the Gaussian significance 
\begin{align}
\# \sigma = \frac{S}{\sqrt{B + \alpha^2 B^2 + \beta^2 S^2}} \; ,
\end{align}
where $S$ and $B$ are the number of signal and background events
passing the cuts assuming $15~\iab$ of data. The systematic
uncertainties on the background and signal are conservatively given as
$\alpha=20\%$ and $\beta =
10\%$~\cite{Low:2014cba,Cirelli:2014dsa}. As we are scanning over a
range of model parameter space with different characteristics, there
is no good way to display a band of significances for the $20-500\%$
backgrounds. Instead, we will only quote the central background
estimate. The left panel of Figure~\ref{fig:RunPoints} shows the
representative sample of points that we used mapped on the surface as
well as the calculated significance. It appears that most of the wino
plateau is covered and that the search works better for larger values
of $|\mu|$.\bigskip

\begin{figure}[t]
\includegraphics[scale=.75]{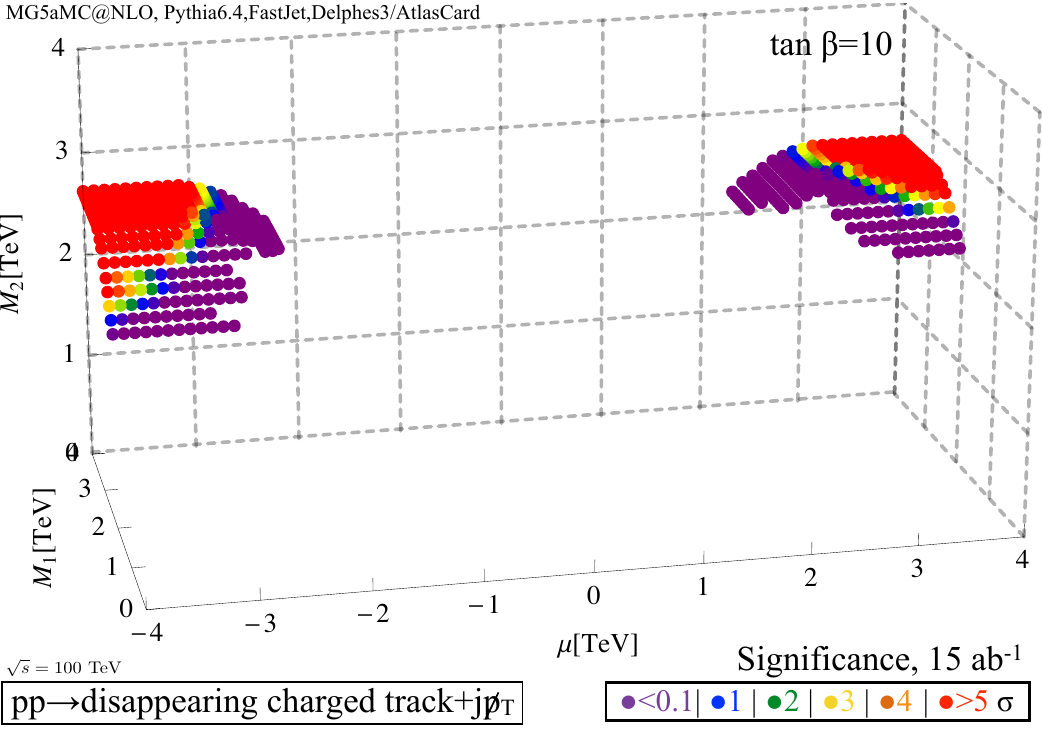} 
\includegraphics[scale=.75]{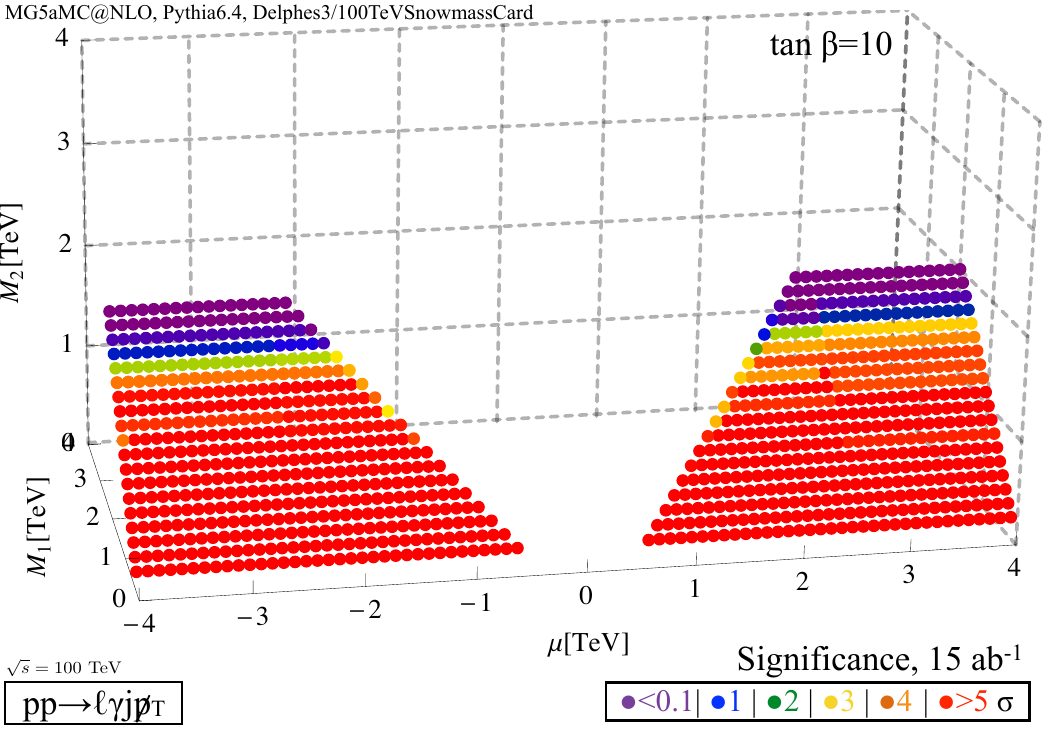}
\caption{\textbf{Left panel:} Points on the relic neutralino surface,
  which will be excluded or discovered using a disappearing track
  search with $15~\iab$ at a 100~TeV collider. At smaller values of
  $|\mu|$ the higgsino still mixes enough to cause the mass splitting
  of the wino plateau to be too large for the disappearing track
  search to be effective. \textbf{Right panel:} Points which will be
  excluded or discovered using a compressed search for $pp \rightarrow
  \ell^\pm \gamma j \met$.}
\label{fig:RunPoints}
\end{figure}

For the points on the relic neutralino surface,
if the decay length is less than 15~mm, the
charginos have almost no chance of traveling far enough to be
registered as a track. We find that for tracks longer than this, at
least in the range we are considering, the points can be fit well by
a cubic function. We focus on the relic neutralino points with a mass
difference between the chargino and the neutralino smaller than
$0.5~\gev$ and find their significance based on the best fit cubic
curve. We then plot the points that can be discovered at $5\sigma$ and
those which can be excluded at $2\sigma$.  The result is shown in
Figure~\ref{fig:RunPoints}. We see that most of the wino plateau is
within reach, but as mixing with bino and higgsinos grows, so does the
chargino-neutralino mass splitting. The chargino decay length then
decreases, making the search less effective.


\subsection{Compressed search}

Our compressed bino-wino search is directed at neutralinos with mass
eigenstates separated by $1-40~\gev$ and follows the previous study of
Ref.~\cite{Bramante:2014tba}. It targets events with missing
transverse momentum, photons, and leptons emitted in the decay of
heavier neutralinos. The dominant production and decay process on the
relic neutralino surface is
\begin{align}
pp \rightarrow (\lsp_2^0\rightarrow \gamma \lsp_1^0) \; 
               (\lsp_1^\pm \rightarrow \ell^\pm \nu_\ell \lsp_1^0 ) j 
   \rightarrow \lsp_1^0 \lsp_1^0 \ell^\pm \nu_\ell \gamma j \; , 
\label{eq:comprod}
\end{align}
where the one-loop radiative decay of $\lsp_2^0$ will be more probable
as the neutralino mass splitting decreases.\bigskip

As noted in the introduction to this section, for $M_2 \lesssim
2~\tev$, thermal relic neutralino mass states are arranged so that a
wino-like NLSP is $10-40~\gev$ heavier than a bino-like LSP. This
electroweakino spectrum is especially amenable to searches at a
100~TeV proton-proton collider, because the lepton and photon in the
dominant Standard Model background process $pp \rightarrow W^\pm
\gamma j \rightarrow \nu_\ell \ell^\pm \gamma j$ tend to have higher
transverse momenta whenever the final state neutrino carries enough
momentum to fulfill a hard $\met \gtrsim~\tev$ requirement. The cuts
we employ in this study are
\begin{align}
p_{T,\ell} &=[10-60]~\gev
 & |\eta_\ell| &< 2.5 \notag \\
p_{T,\gamma} &=[10-60]~\gev
 & |\eta_\gamma| &<2.5   & \Delta R_{\ell \gamma} &> 0.5 \notag \\
p_{T,j} &> 0.8~\tev  & |\eta_j| &<2.5 & M_{T2}^{(\gamma,\ell)} &< 10~\gev \notag \\
\met &> 1.2~\tev  \; .
\label{eq:compcuts}
\end{align}

The cut on the lepton-photon separation, $\Delta R_{\ell \gamma}$, reduces
background events in which the lepton or $W^\pm$ radiates a
photon. The upper limit on the stransverse
mass~\cite{Lester:1999tx,Barr:2003rg,Lester:2007fq,Cheng:2008hk} of
the lepton and photon, $M_{T2}^{(\gamma,\ell)}$, removes $W \gamma j$
background events: in these events the photon direction is less
correlated with $\met$ than for a decaying neutralino, $\lsp_2^0
\rightarrow \gamma \lsp_1^0$. We specifically use the bisection-based asymmetric
$M_{T2}$ algorithm of \cite{Lester:2014yga}. To reject hadronic backgrounds, events
with more than two jets with $p_{T,j} > 300~\gev$ are vetoed. To
reject electroweak backgrounds, events with more than one lepton or
photon are rejected. For a lengthy discussion of this search,
including the effect of background events with jets faking photons,
see Ref.~\cite{Bramante:2014tba}.\bigskip

In the right panel of Figure~\ref{fig:RunPoints} we show the
significance attained, assuming 5\% signal and background uncertainty
($\alpha=\beta=0.05$), after $15~\iab$ luminosity at a 100~TeV
collider, obtained by simulating the signal given in
Eq.\eqref{eq:comprod} with the dominant $W \gamma j$ background.  In
this collider study, supersymmetric masses are set with
\toolfont{SuSpect}~\cite{Djouadi:2002ze} (without loop corrections,
but with inter-chargino-neutralino mass splittings manually determined
using loop-level custodial symmetry breaking mass splittings, as
described in Section~\ref{sec:sommersurface}). The decay widths are
computed with \toolfont{SUSY-HIT}~\cite{Djouadi:2006bz}, and events
are generated at tree-level in
\toolfont{MG5aMC@NLO}~\cite{Alwall:2014hca} and
\toolfont{Pythia6.4}~\cite{Sjostrand:2006za}. Jets are clustered using
the anti-$k_T$ algorithm~\cite{Cacciari:2008gp} in
\toolfont{Delphes3}~\cite{deFavereau:2013fsa}, with the Snowmass
100~TeV detector card introduced in Ref.~\cite{Avetisyan:2013onh}.

\begin{figure}[t]
\includegraphics[scale=.8]{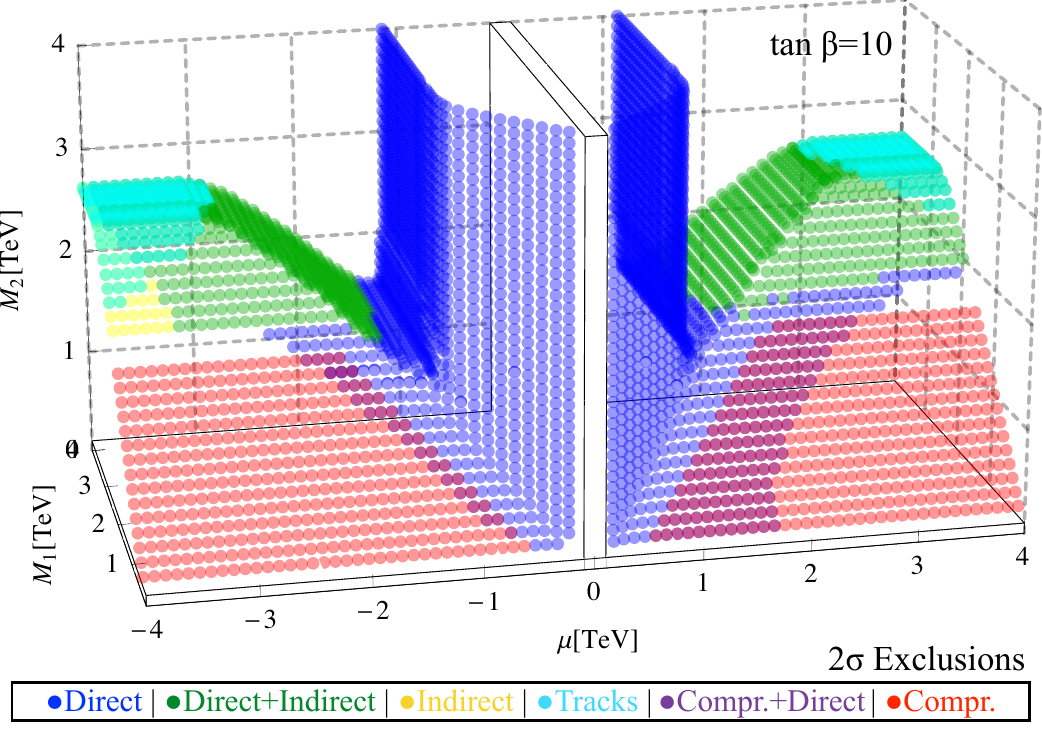}
\caption{A combination of $2 \sigma$ exclusions from future indirect
  (CTA and HAWC), direct (XENON1T and LZ), and collider searches
  (charged tracks and compressed events at 100~TeV) are shown over the
  surface of thermal relic neutralinos.}
\label{fig:moneyplot}
\end{figure}

\section{Conclusions}
\label{sec:conclusion}

We have systematically studied the phenomenology of the thermal relic
neutralino dark matter surface, incorporating the effect of
Sommerfeld-enhancement in setting the relic abundance at neutralino
freeze-out. Spin-independent direct detection experiments will explore
much of the relevant parameter space, including that of nearly-pure
higgsino LSP, so long as $M_1,M_2 < 4~\tev$. Regions of nearly-pure
wino LSP with will be probed by future galactic center gamma ray
searches, and also with charged track searches at a future 100~TeV
hadron collider. Regions with a bino-like LSP, and particularly the
bino-wino space with $M_{1,2} < 2~\tev$ and $|\mu| \gtrsim 1.5 ~\tev$
can only be accessed with future compressed electroweakino searches at
a 100~TeV collider (or a $\sqrt{s}\geq 4~\tev$ electron-positron
machine~\cite{Aicheler:2012bya}). We plot $2\sigma$
exclusions of different futures experiments in
Figure~\ref{fig:moneyplot}, finding a solid coverage of the sommerfelded
thermal relic neutralino surface.

\acknowledgments We thank Andrzsej Hryczuk for useful correspondence
and the use of \toolfont{DarkSE}. We would also like to thank Zhenyu
Han and Graham Kribs for useful discussions. JB thanks CETUP 
for hospitality and support, as well 
as the Aspen Center for Physics, which is supported by 
National Science Foundation grant PHY-1066293. ND was partially supported by the Alexander von Humboldt Foundation. BO was partially supported by the U.~S.~Department of Energy under Grant Nos. DE-SC0011640. The work of AM was partially supported by the National Science Foundation under Grant No. PHY-1417118. We specifically acknowledge the assistance of the Notre Dame Center for Research Computing for computing resources.

\bibliography{serelic}

\end{document}